\documentclass[fleqn,10pt]{wlscirep}

\usepackage{graphicx}
\usepackage{subcaption}
\usepackage{amssymb}
\usepackage{amsmath}
\usepackage{url}
\usepackage{hyperref}
\usepackage{algorithm}
\usepackage{algorithmic}
\usepackage{multirow}

\usepackage{todonotes}

\graphicspath{ {figures/} }

\title{Priority Attachment: a Comprehensive Mechanism for Generating Networks}

\author[1*]{Mikołaj Morzy}
\author[2]{Tomasz Kajdanowicz}
\author[2]{Przemysław Kazienko}
\author[1]{Grzegorz Miebs}
\author[1]{Arkadiusz Rusin}
\affil[1]{Institute of Computing Science, Poznań University of Technology, 60-965 Poznań, Poland}
\affil[2]{ENGINE - The European Centre for Data Science, Faculty of Computer Science and Management, Wrocław University of Science and Technology, 50-370 Wrocław, Poland}

\affil[*]{Mikolaj.Morzy@put.poznan.pl}

\begin{abstract}

We claim that networks are created according to the priority attachment mechanism. We introduce a simple model, which uses the priority attachment to generate both synthetic and close to empirical networks. Priority attachment is a mechanism, which generalizes previously proposed mechanisms, such as small world creation or preferential attachment, but we also observe its presence in a range of real-world networks. In this paper, we show that by using priority attachment we can generate networks of very diverse topologies, as well as re-create empirical ones. An additional advantage of the priority attachment mechanism is an easy interpretation of the latent processes of network formation. We substantiate our claims by performing numerical experiments on both synthetic and empirical networks. The two main contributions of the paper are: the development of the priority attachment mechanism, and the design of the Priority Rank: a simple network generative model based on the priority attachment mechanism. 

\end{abstract}

\begin{document}

\flushbottom
\maketitle

\thispagestyle{empty}

\section*{Introduction}
\label{sec:introduction}

\subsection*{Motivation}
\label{subsec:motivation}

Many generative models of network formation have been proposed in the scientific literature~\cite{chakrabarti2006graph}, and some of them have gained significant notoriety, for instance the random network model of Erd\"os and R\'enyi~\cite{erd6s1960evolution}, the small world model of Watts and Strogatz~\cite{watts1998collective}, the cumulative advantage model of de Solla Price~\cite{price1976general}, the scale-free model of Albert and Barab\'asi~\cite{barabasi1999emergence}, stochastic block model~\cite{BlockModel_First_Steps}, \cite{wang1987stochastic} or the forest fire model of Leskovec~\cite{leskovec2005graphs}. Each of these generative network models is based on some phenomenon which (as is often claimed) explains the underlying process of network formation. For instance, in the case of the small-world model, the alleged phenomenon is the tendency of many systems to form tightly connected groups (small worlds) with incidental connections between groups serving as long distance bridges. In the case of the preferential attachment model the phenomenon which purportedly fuels the network formation process is the strong preference of vertices to connect to already well-connected vertices. In the stochastic block model, which is is a probabilistic model of the interactions between pairs of vertices, each vertex belongs to one of the groups or 'blocks', and each edge exists with a certain probability depending only on the vertices' membership to the particular group. Fluctuations in the same block are stochastically equivalent, indicating their equivalent role in generating of the network structure. This model is popular because it can generate a wide range of large-scale networking patterns and can learn social structure in unweighted and weighted  networks~\cite{BlockModel_weighted}, \cite{BlockModel_structure_learning},\cite{guimera2009stochastic}.

Some of the network generators do not attempt to model real-world processes directly, but they rely on some mathematical formalism, like the sequence of Kronecker products applied to a small seed set of networks~\cite{leskovec2010kronecker}. Recently, an interesting proposal has been formulated to model complex networks using a stochastic sequence of predefined base actions~\cite{arora2017action} and turning the re-creation of an empirical network into an optimization problem. 

For a long time, we have suspected that these individual phenomena are specialized instances of a more general mechanism of network creation. The main reason for this forefeeling was the fact that the generative network models seemed to be narrowly defined and each of them covered only a specific class of possible network topologies. A closer inspection of the generative network models further revealed that all of them were using, sometimes inadvertently, some type of prioritization while choosing target vertices during edge formation. The mechanism of prioritization using priority queues has a very long presence in almost all disciplines of science. In psychology priority queues are believed to be the driving mechanisms of human attention~\cite{yantis1990mechanisms}. In systems science priority queues play an important role in controlling and scheduling~\cite{cobham1954priority, moon2000scalable} or in recommender systems for evolving networks\cite{liao2017ranking}. In sociology there are strong clues that the stochastic queuing theory can provide valuable insights into human dynamics~\cite{walraevens2012stochastic, jo2012time, alexander2012priority}. In ecology prioritization influences the inter-species interactions~\cite{louette2007predation, almany2003priority, blaustein1996priority}.

In the domain of complex networks, the idea of network growth based on rankings was introduced by Fortunato, Flammini, and Menczer\cite{fortunato2006scale}. They have shown that substituting vertex feature distributions with global rankings\footnote{Within the scope of this paper we consider \emph{global ranking} to be a universal ordering of vertices common to all vertices, whereas \emph{local ranking} denotes an ordering of vertices as perceived by an individual vertex.} of these features (both topological and non-topological) as the basis for preferential attachment, one always obtains a scale-free degree distribution in the resulting network. In addition, they have proven that it is sufficient to use ranking samples in the network growth process, as long as these samples maintain the original partial order of vertices. However, the approach presented in this paper differs from \cite{fortunato2006scale} in that we consider not only global rankings of vertices, but we allow local rankings as well. This is a fundamental difference since our model is not limited to scale-free networks, but can produce arbitrary network topologies. In addition, our model provides means for computational re-creation of empirical networks resulting in many instances of networks with nearly identical network profiles. It is a unique and promising property in the field of complex networks.

In order to verify whether the priority attachment can explain processes of network formation, we have developed Priority Rank, a simple generative network model. This model uses the priority attachment phenomenon to drive the network formation. Our experiments show that Priority Rank is capable of generating a very wide spectrum of networks. We were able to successfully re-create network topologies generated by the most popular generative network models, which suggests that the priority attachment mechanism is indeed the common denominator for these models. More importantly, examining the prioritization scheme, which resulted in the best re-generation of the original network, we could provide viable interpretations of the latent network generative process. This unique feature of the Priority Rank model is best pronounced in empirical networks.

\subsection*{Problems addressed}
\label{subsec:problems.addressed.in.the.paper}

Research on complex networks is constantly struggling with a serious methodological obstacle. Despite the vast availability of networks that can be analyzed, each individual network has its distinctive characteristics and presents to a researcher a single data point. Any attempt to generalize the results obtained on one network to other networks strives with inherent vulnerabilities. The choice of target networks is arbitrary and the transfer of discovered patterns has to be selective. Oftentimes, unspoken assumptions come into play, for instance, a phenomenon discovered in a few networks of a particular class is presumed to be applicable to all networks of this class. These assumptions are much more questionable than usually assumed and they significantly weaken the scientific appeal of a discovery. Let us consider the phenomenon of the shrinking diameter of a network~\cite{leskovec2005graphs}. Is it fair to conclude that this phenomenon should apply to all complex networks? Or maybe it is valid only for scale-free networks characterized by the preferential attachment mechanism of edge formation? Unfortunately, it is impossible to conduct a proper statistical inference since only one realization of each network is available. Furthermore, this problem cannot be addressed by sampling of networks, because sampling distorts the topology of networks to the point where sub-networks of a network loose their identifying properties~\cite{Stumpf22032005}.

Another problem addressed in this paper is the inherent incompatibility of networks. Trying to analyze different networks and to make quantitative comparisons between them is very difficult, because in the realm of complex networks even minuscule perturbations of local topology can lead to significant differences in global network properties. This is true even for synthetic networks which have been generated from the same generative model and with slightly different parameters. Distributions of centrality measures, such as degree, betweenness, closeness or clustering coefficient, vary dramatically in response to minor changes of initial configurations~\cite{e18090320}.  As the result, statistical comparison of seemingly similar networks suggests a much greater variability and difference between networks, which makes any machine learning extremely difficult in the domain of complex networks. Without the access to diversified training, testing, and validation sets of networks, all patterns discovered in networks are bound to overfit and not generalize well. Also, the relationships between network generation processes proposed in the literature are very unclear. Is the preferential attachment an alternative to the small world phenomenon, or is it merely a supplementation? We note that there is no common language at the level of network generating processes, making the comparisons between different networks even more challenging.

\subsection*{Main findings}
\label{subsec:main.results}

The main contribution presented in this paper is the priority attachment mechanism of network formation. We show how priority queues built on local rankings can lead to very diversified network topologies. The function, which generates these local rankings, yields itself to interpretation. As it turns out, popular generative network models can be easily mimicked by the priority attachment, and some of these models (preferential attachment, cumulative advantage) are a special case of the priority attachment. The phenomenon of the priority attachment can be readily incorporated into a generative Priority Rank network model, which is a simple procedure of network creation. The Priority Rank model can produce networks with very different topologies using the same principle of priority attachment. Instead of using dedicated network models one can successfully force Priority Rank to produce random networks, preferential attachment networks, small world networks, and many more. The second main feature of Priority Rank is its applicability to empirical networks. Due to the priority attachment mechanism, which forms the basis of the model, it is possible to adjust Priority Rank to deliver networks with topology very similar to a given empirical network. The Priority Rank model has only to learn the distance function used in the priority attachment. 
We demonstrate Priority Rank's ability to re-create empirical networks while preserving the distributions of centrality measures and other network characteristics. This ability allows for generating whole families of similar networks, thus, for conducting statistical inference on families of networks, etc.


There are four main reasons, which make the Priority Rank model valuable:

\begin{itemize}
    \item Priority Rank is a comprehensive generative network model which can produce a very wide spectrum of network topologies, including the most popular network models.
    \item Priority Rank offers insights into the generative processes behind modeled networks. Classical machine learning algorithms can be used to find the most fitting distance function for the priority attachment mechanism. In most cases such a distance function is easily interpretable and provides explanations for the latent network formation process.
    \item Priority Rank allows us to generate multiple instances of networks with the same characteristics and distributions, because the model discovers the main generative process of network formation. It should be noted that this is incomparable with network sampling which oftentimes distorts the profile of sampled networks. Instead, the Priority Rank model allows to multiply networks for the purpose of A/B testing, statistical inference, simulations, etc.
    \item Priority Rank does not require any hyper-parameters to be set \emph{a priori} such as edge creation probability in the random network model or edge rewiring probability in the small world network. This feature of the Priority Rank model is very important because, contrary to popular belief, generative network models are very sensitive to the initialization of these parameters.
\end{itemize}

\section*{Methods}
\label{sec:methods}

\subsection*{Priority attachment}
\label{subsec:priority.attachment}

The idea behind the priority attachment mechanism is fairly simple. Consider a new vertex which joins a network. The primary issue is the selection of target vertices to which the new vertex creates edges. Previously, several mechanisms have been proposed to model this selection process. For instance, in the Erd\"os-R\'enyi model the vertex selects target vertices randomly using the uniform distribution. In the Albert-Barab\'asi model the vertex selects target vertices with the probability proportional to the current degree of each vertex. According to the priority attachment mechanism each vertex has a local ranking which arranges all possible target vertices by their ``importance'' from the point of view of the new vertex. The new vertex selects target vertices with the probability proportional to their \emph{position in the local ranking}. One should regard this local ranking as the priority queue which orders all of the vertices of the network from the point of view of a single vertex. This means that each vertex uses its own local ranking while creating edges. In other words, the main mechanism of network formation is the attachment of vertices driven by their individual perception of priority of other vertices, hence the name "priority attachment". The power of the priority attachment mechanism stems from the fact that local rankings can be computed by arbitrarily complex distance functions which can either model real-world phenomena, or model adjacency matrices of empirical networks.

\begin{figure}[ht]
\centering
\begin{subfigure}[b]{0.45\textwidth}
    \includegraphics[width=\textwidth]{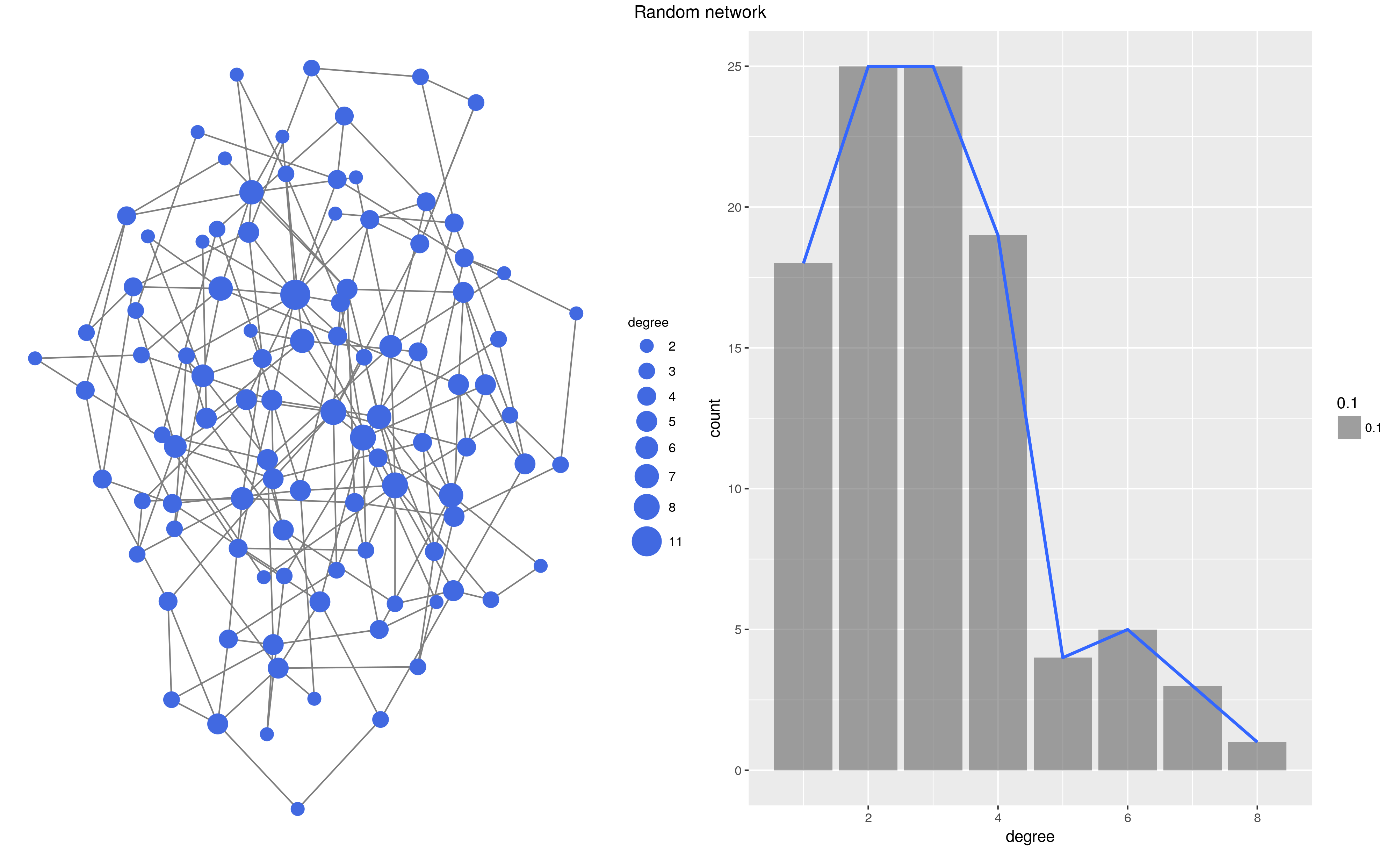}
    \caption{Random network}
\end{subfigure}
\begin{subfigure}[b]{0.45\textwidth}
    \includegraphics[width=\textwidth]{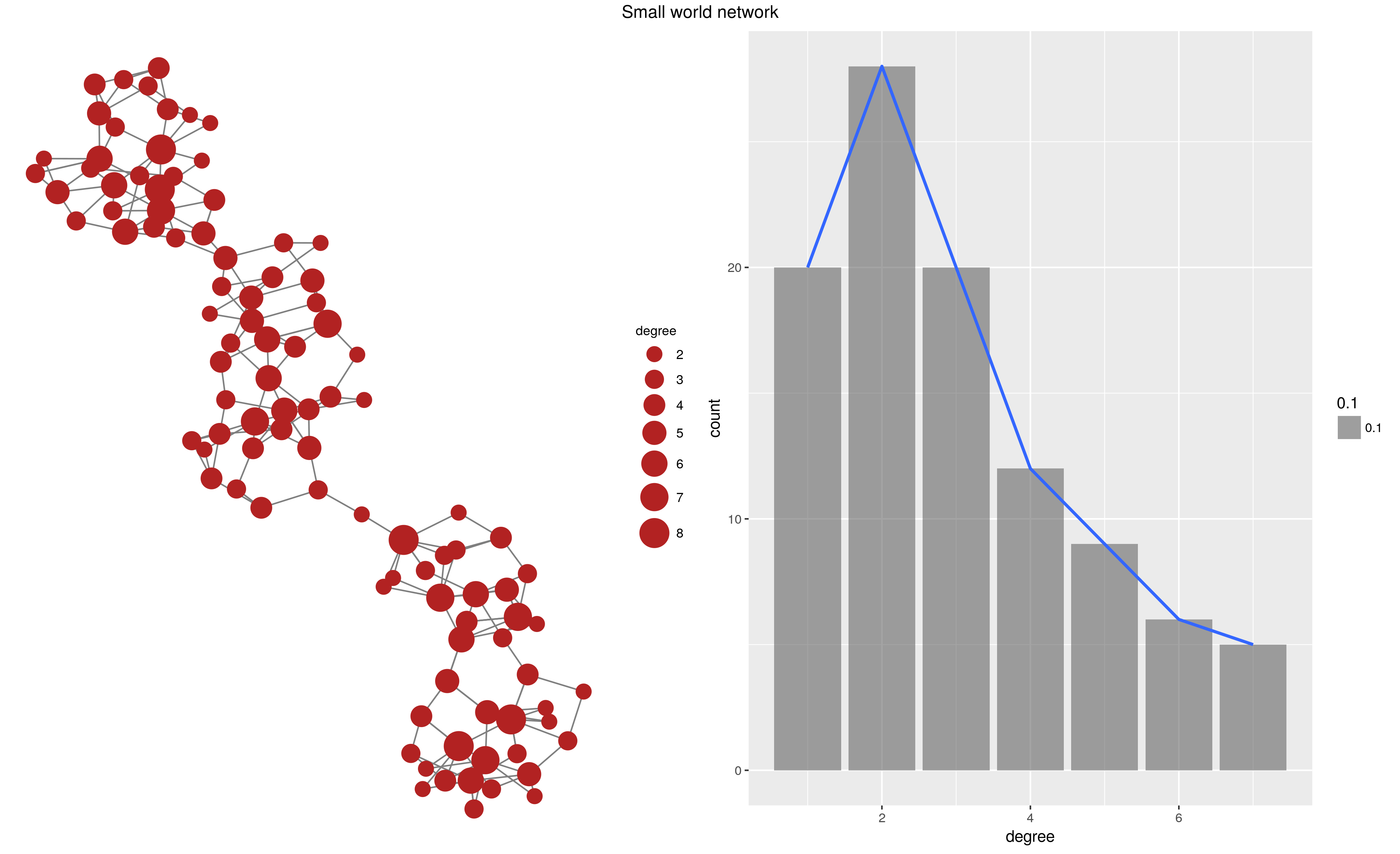}
    \caption{Small-world network}
\end{subfigure}
\begin{subfigure}[b]{0.45\textwidth}
    \includegraphics[width=\textwidth]{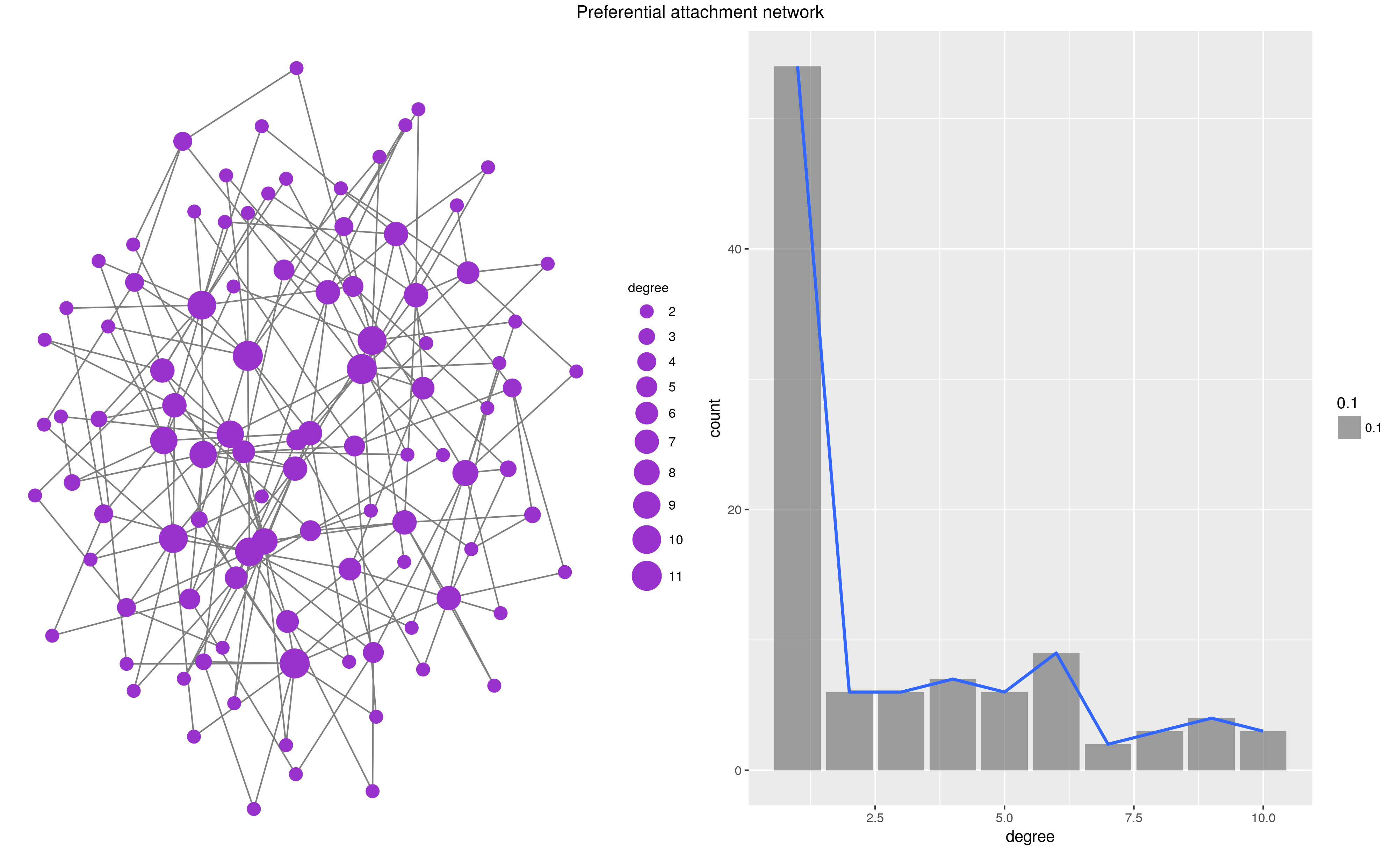}
    \caption{Preferential attachment network}
\end{subfigure}
\begin{subfigure}[b]{0.45\textwidth}
    \includegraphics[width=\textwidth]{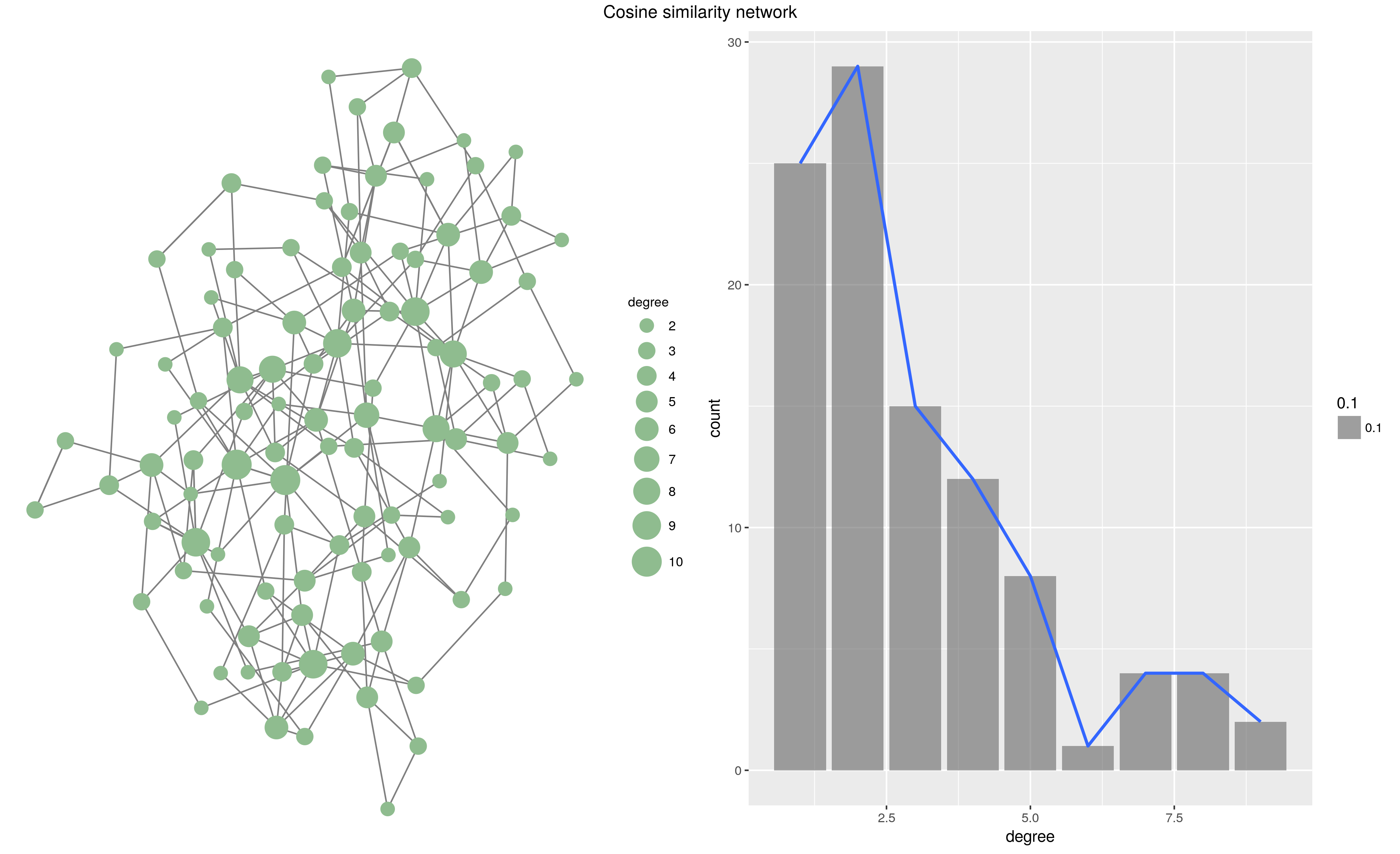}
    \caption{Cosine similarity network}
\end{subfigure}
\begin{subfigure}[b]{0.45\textwidth}
    \includegraphics[width=\textwidth]{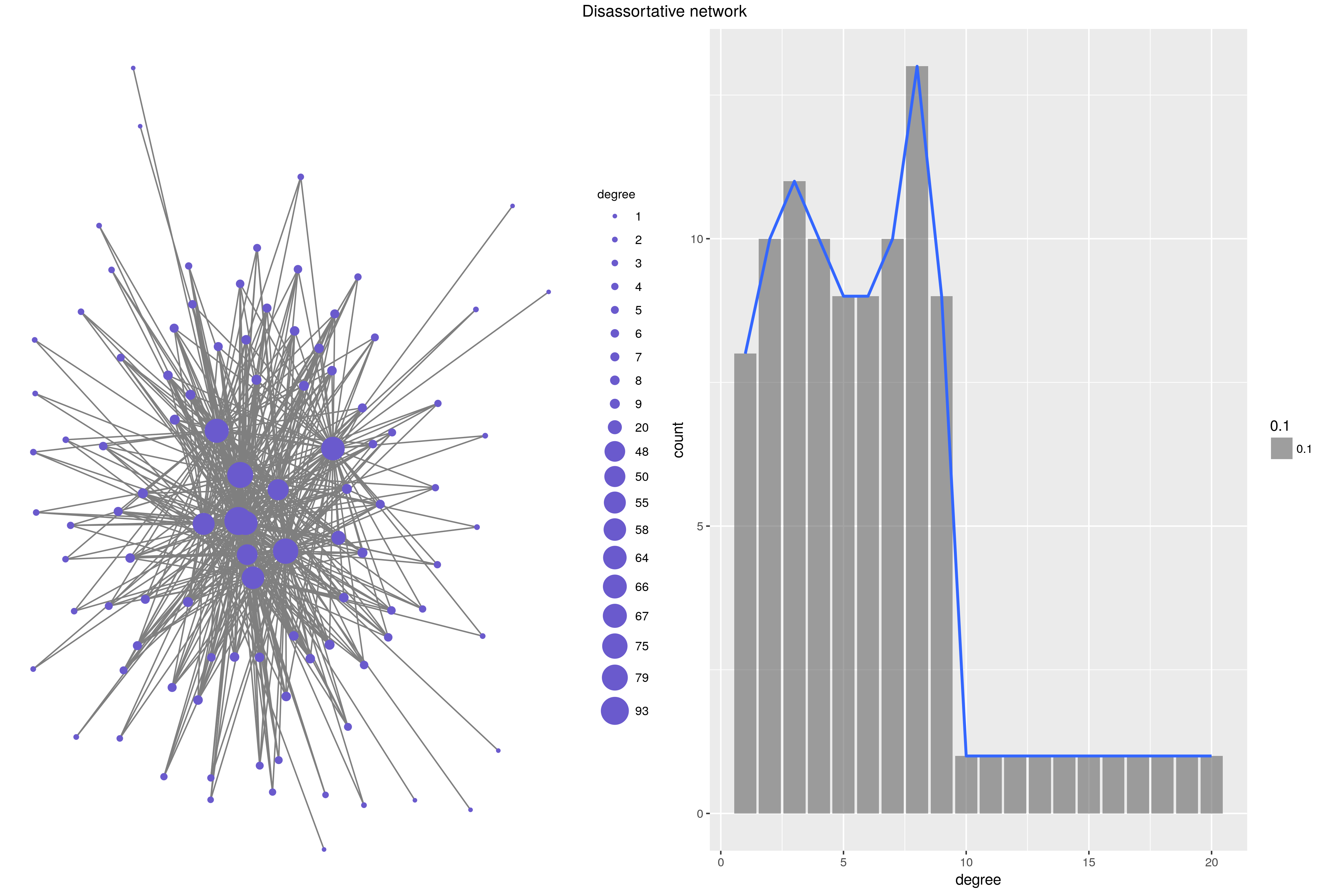}
    \caption{Disassortative network}
\end{subfigure}
\begin{subfigure}[b]{0.45\textwidth}
    \includegraphics[width=\textwidth]{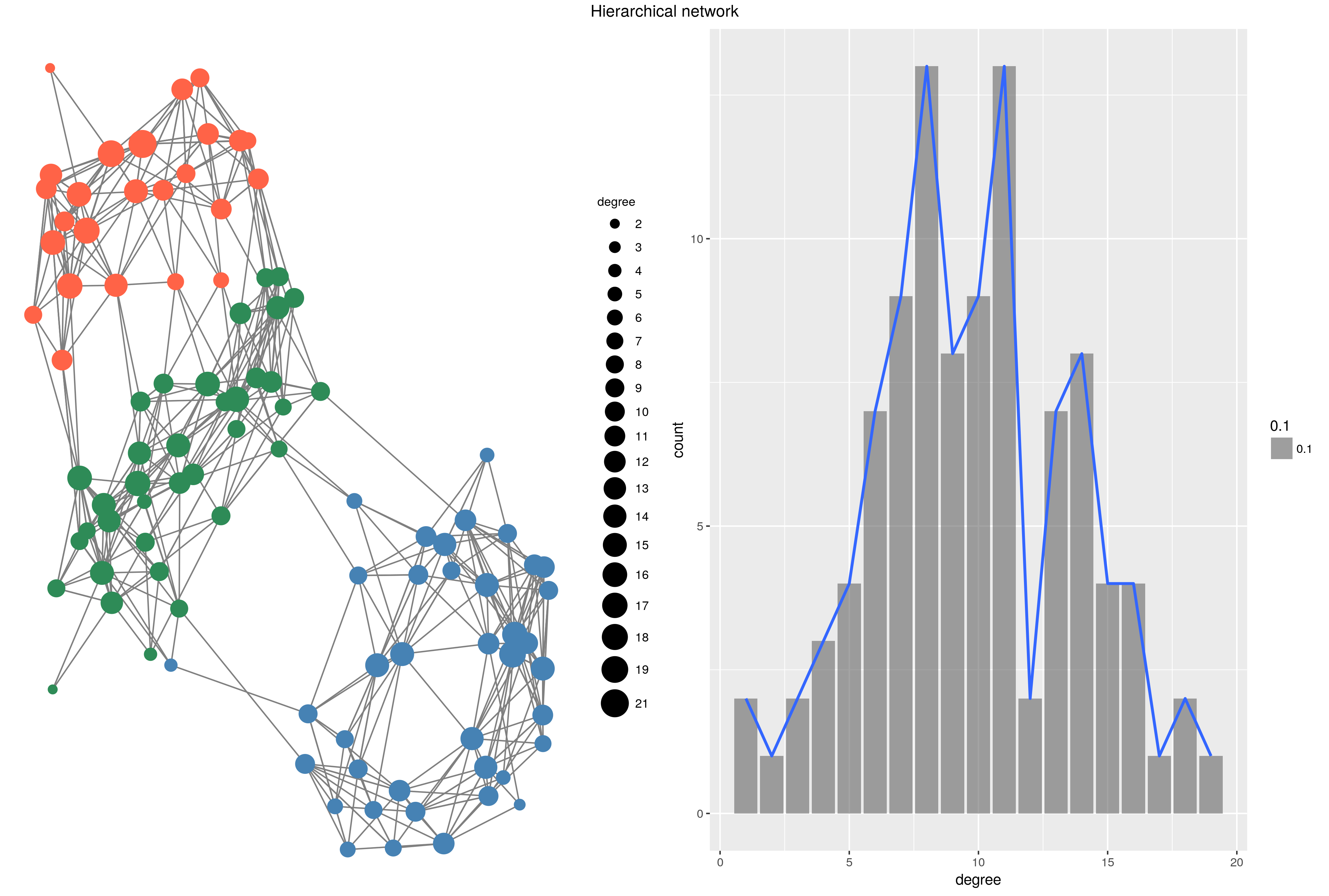}
    \caption{Hierarchical network}
\end{subfigure}
\caption{Different network topologies and their degree distributions generated by the priority attachment mechanism}
\label{fig:network.topologies}
\end{figure}

Interestingly, the topology of the generated network depends almost exclusively on the properties of the function $D$ which is used to generate local rankings. Figure~\ref{fig:network.topologies} presents six different networks, each consisting of $n=50$ vertices, generated by the priority attachment mechanism. Random network is generated for $D(v_i,v_j) \sim N(\mu, \sigma)$, i.e., when local rankings are random permutations of the set of vertices. Small-world network is generated for $D(v_i,v_j) = \left| a(v_i) - a(v_j) \right|$ for an attribute which value is randomly chosen from a uniform distribution, $a(v) \sim U(0,1)$. In other words, local rankings arrange vertices by the distance defined by the vertex's attribute $a(v)$, resulting in a strong preference for vertices in the local neighborhood. 
Preferential attachment network is generated for $D(v_i,v_j) = \frac{1}{\mathit{deg}(v_j) + \epsilon}$, i.e., when local rankings simply represent the global ranking of vertex degrees\cite{fortunato2006scale}. Cosine similarity network is generated if $D(v_i,v_j) = 1 - \frac{v_i \circ v_j}{\|v_i\|\|v_j\|}$, where $v_i=\left[a^1_i, a^2_i,\ldots,a^m_i\right]$ is a vector of numeric values, effectively linking vertices which are close in the vertex vector space. 

But priority attachment mechanism can be used to produce networks with more complex topologies. For instance, the function $D(v_i,v_j) = \mathit{deg}^* - \frac{\mathit{deg}(v_i)}{\mathit{deg}(v_j)}$, where $\mathit{deg}^*$ is the maximum degree in a given network, produces a regular disassortative network with the degree assortativity coefficient $r = -0.28$. Another example is the function $D(v_i,v_j) = \alpha D_E(v_i,v_j) + (1-\alpha) D_H(v_i, v_j)$, where $D_E(v_i, v_j)$ is the Euclidean distance between vertices $v_i$, $v_j$ and $D_H(v_i, v_j)$ is the hierarchical distance between these two vertices (we assume that vertices belong to disjoint classes and that there is a total ordering between classes). When applied to the Priority Rank model, such distance function produces complex hierarchical structures. As can be seen in Figure~\ref{fig:network.topologies}, networks created upon various functions $D$ have different topologies and different degree distributions.

The main advantage of these simple distance function definitions is the fact that they are easily interpretable. For instance, when modeling a network of disease spreading it is likely that disease vectors would infect vertices in their close physical proximity. Thus, a distance function based on the physical distance between vertices (which generates small-world structures) would be a sound choice for a simple model of disease spreading. In addition, if one would like to differentiate the probability of edge formation based on additional factors (e.g., the transfer of a sexually transmitted disease is more likely between vectors of similar age), the incorporation of this factor into the distance function would be trivial. Similarly, when trying to model semantic relationships between words embedded in multidimensional space (a standard tool in the contemporary NLP), it is reasonable to assume that the proximity of word embeddings is an indication of some semantic relatedness of the words. A simple way to model these relationships would be to use cosine distance to define the distance function $D$.

Let us now formally define the priority attachment mechanism and present the Priority Rank model. Further, we will use the following notation.

\begin{itemize}
    \item $G = \left< V, E \right>$ is a network with the set of $n$ vertices $V = \{ v_1, \ldots, v_n \}$ and the set of edges $E = \{ (v_i, v_j) : v_i, v_j \in V ; i \neq j\}$
    \item $\forall v_i \in V : v_i = \left[ a^1_i, a^2_i, \ldots, a^m_i \right]$, vertices are vectors of attributes,
    \item $D(v_i, v_j): V \times V \rightarrow \mathbb{R}$ is the generic distance function which computes the distance between vertices $v_i$ and $v_j$, such that $D(v_i,v_j) > 0 \iff i \neq j$ and $D(v_i,v_i)=0$. Distance function $D$ does not have to be symmetrical.
    \item $R_i = \left< v_i^1, v_i^2, \ldots, v_i^{n-1} \right>$ is a permutation of $V \setminus \{v_i\}$ representing the local ranking of vertices for the vertex $v_i$
\end{itemize}



According to the priority attachment mechanism, the probability of selecting a vertex $v_j$ as the target vertex for an edge originating from the vertex $v_i$ is inversely proportional to the position of the vertex $v_j$ in the ranking of vertices for $v_i$. The probability mass function of selecting the \emph{i}th element of the ranking is given by

\begin{equation}
    P(i) = \frac{1}{\sum\limits_{k=1}^{n-1} \frac{1}{k}} \frac{1}{i} = \frac{1}{H_{n-1} i}
\label{eq:pmf}
\end{equation}

where $H_n$ is the \emph{n}th harmonic number, serving as the normalizing constant so that Equation~\ref{eq:pmf} presents a proper probability mass function, i.e., $\sum_{i=1}^n P(i) = 1$. We will use Euler's formula to approximate the \emph{n}th harmonic number as $H_n \approx ln(n) + \frac{1}{2n} + \gamma$, where $\gamma$ is the Euler-Mascheroni constant, $\gamma = 0.57722$. Algorithm~\ref{alg:network.generation} presents the pseudo-code for generating networks using the Priority Rank model. Procedure $\mathit{sample}(\left[1,\ldots,m\right],P)$ samples an integer from the range $\langle 1,\ldots,m \rangle$ without replacement using the probability mass function from Equation~\ref{eq:pmf}. We assume a constant out-degree $k$ of vertices, but the number of edges each vertex creates can vary from vertex to vertex. In particular, when using Priority Rank model to re-create empirical networks, one can sample the out-degree distribution of the source network to obtain a particular value of $k$ for a given vertex.

\begin{algorithm}[ht]
    \begin{algorithmic}[1]
    \REQUIRE $D : V \times V \rightarrow \mathbb{R}$, distance function
    \REQUIRE $V={v_1,\ldots,v_n}$, a set of vertices
    \REQUIRE $k$, the number of edges each vertex creates
    \FOR{$i=1$ to $n$}
        \STATE compute the ranking $R_i$ using the distance function $D$  \hspace{1cm} /* \emph{priority evaluation} */
        \FOR{$j=1$ to $k$}
            \STATE $t \leftarrow \mathit{sample}(\langle 1,\ldots,n \rangle, P)$ \hspace{1cm} /* \emph{random selection without replacement of the position in the ranking} */
            \STATE $v_t \leftarrow R_i[t]$ \hspace{3.2cm} /* \emph{priority attachment} */
            \STATE add edge $(v_i,v_t)$
        \ENDFOR
    \ENDFOR
\end{algorithmic}
\caption{Priority Rank generative network model}
\label{alg:network.generation}
\end{algorithm}

In order to better illustrate the idea of priority attachment, let us consider an example of a simple network formation. Suppose that there are five people $v_1,..,v_5$ described by \textit{name, age}, and \textit{sex}. Let us also assume that the distance function $D$ is defined as follows: 

\vspace{0.5cm}

\begin{minipage}[b]{0.35\linewidth}
    \centering
    \begin{tabular}{cccc}
    \hline
      & name  & age & sex    \\ \hline
      $v_1$ & Alice & 30  & female \\
      $v_2$ & Bob   & 40  & male   \\
      $v_3$ & Cecil & 25  & male   \\ 
      $v_4$ & Diana & 20  & female \\
      $v_5$ & Eve   & 35  & female  
    \end{tabular}
\end{minipage}
\hspace{0.5cm}
\begin{minipage}[b]{0.65\linewidth}
    \centering
    \begin{equation*}
        D(v_i,v_j) = 
        \begin{cases}
            |v_i[\mathit{age}]-v_j[\mathit{age}]| ,& \text{if } v_i[\mathit{sex}] = v_j[\mathit{sex}] \\
            |v_i[\mathit{age}]-v_j[\mathit{age}]| + 10,& \text{otherwise}
        \end{cases}
    \end{equation*}
\end{minipage}

\begin{figure}[ht]
\centering
    \includegraphics[width=\textwidth]{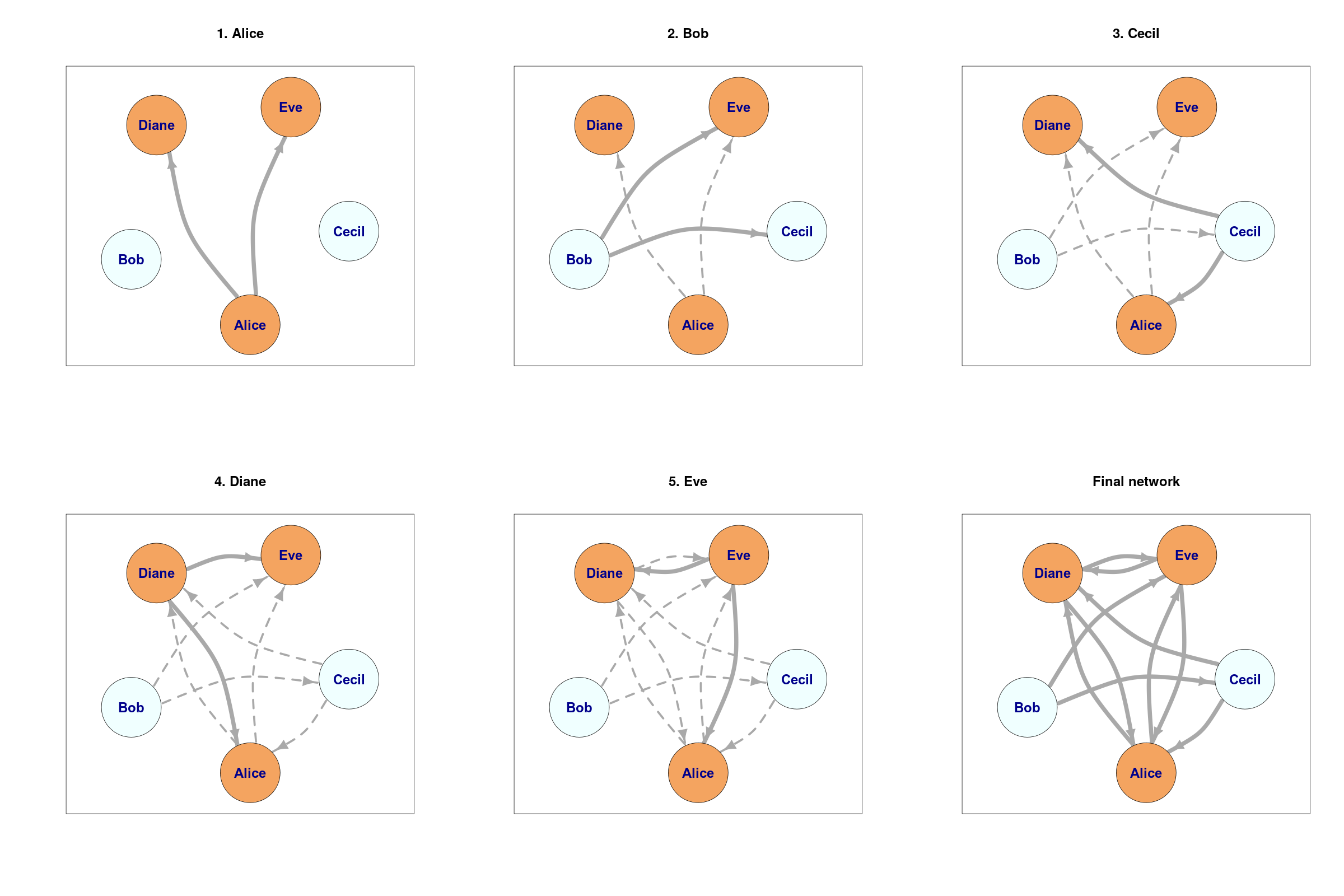}
\caption{Priority attachment process. Edges created at a given step, i.e., by a given person, are marked with solid lines.}
\label{fig:priority.attachment}
\end{figure}

In other words, the social distance is defined in terms of the absolute difference of age, and the fact that two people share the same sex compensates for 10 years of age difference. In this example, people tend to form relationships with other people of similar age, and given two people of the same age, there is a preference to form relationships with people of the same sex. This model could be applicable, for instance, to the process of self-selecting students to form pairs in a large study group where the participants have no prior acquaintances. One possible instance of the network formation process driven by the priority attachment is presented in Figure~\ref{fig:priority.attachment}, assuming that each vertex always creates $k=2$ outgoing edges. Individual priority rankings for each vertex, along with the value of the distance function and the probability of creating an edge to a vertex (computed using Eq.~\ref{eq:pmf}), are presented in Table~\ref{tab:example.priority.attachment.rankings}. 

The process starts with Alice calculating her distance to all other vertices. The most similar vertex to Alice is Eve and she occupies the first position in the local ranking for Alice. Analogously, the most dissimilar vertex to Alice is Bob and he is placed at the end of the ranking. Inserting ranking positions into Equation~\ref{eq:pmf} yields the final probabilities of selecting vertices as target vertices for newly created edges. For the sake of simplicity we have assumed that for each vertex the first two most probable targets have been randomly chosen while forming the network. If two or more vertices are equidistant from the given vertex, they receive the same position in the local ranking, which may contain gaps. As the result, probabilities of selecting vertices at certain positions of the local ranking may change (compare local rankings of Alice, Bob, and Cecil).

\begin{table}[ht]
\centering
\begin{tabular}{clcc}
\hline
\multicolumn{4}{c}{\textbf{Alice}}  \\ 
rank & name  & dist & prob \\ \hline
1    & Eve   & 5    & 48\% \\
2    & Diane & 10   & 24\% \\
3    & Cecil & 15   & 16\% \\
4    & Bob   & 20   & 12\%
\end{tabular}
\quad
\begin{tabular}{clcc}
\hline
\multicolumn{4}{c}{\textbf{Bob}}  \\
rank & name  & dist & prob \\ \hline
1    & Cecil & 15   & 39\% \\
1    & Eve   & 15   & 39\% \\
3    & Alice & 20   & 13\% \\
4    & Diane & 30   & 9\%
\end{tabular}
\quad
\begin{tabular}{clcc}
\hline
\multicolumn{4}{c}{\textbf{Cecil}}  \\
rank & name  & dist & prob \\ \hline
1    & Alice & 15   & 31\% \\
1    & Bob   & 15   & 31\% \\
1    & Diane & 15   & 31\% \\
4    & Eve   & 20   & 7\%
\end{tabular}
\\ \vspace{0.5cm}

\begin{tabular}{clcc}
\hline
\multicolumn{4}{c}{\textbf{Diane}}  \\
rank & name  & dist & prob \\ \hline
1    & Alice & 10   & 45\% \\
2    & Cecil & 15   & 22\% \\
2    & Eve   & 15   & 22\% \\
4    & Bob   & 30   & 11\%
\end{tabular}
\quad
\begin{tabular}{clcc}
\hline
\multicolumn{4}{c}{\textbf{Eve}}  \\
rank & name  & dist & prob \\ \hline
1    & Alice & 5    & 45\% \\
2    & Bob   & 15   & 22\% \\
2    & Diane & 15   & 22\% \\
4    & Cecil & 20   & 11\%
\end{tabular}
\caption{Priority attachment local rankings with distances and vertex selection probabilities}
\label{tab:example.priority.attachment.rankings}
\end{table}

A visualization of the Priority Rank model mechanics is available online at \url{https://priorityattachment.ml}.

\subsection*{Re-creating empirical networks}
\label{subsec:recreating.networks}

Given an empirical network $G$, we are interested in finding the distance function $D$ such that this distance function, when used inside the Priority Rank model, generates a network which is ``similar'' to the empirical network $G$. The problem of defining a robust and flexible measure of network similarity has been studied for many years and several network similarity measures have been proposed in the literature~\cite{aliakbary2015distance,schieber2017quantification}. However, these measures tend to be computationally exhaustive and difficult to apply to really large networks. For this reason, we have decided to use a simple and well-understood network similarity measure. In order to measure the degree of network similarity we compare the distributions of centrality measures using the Kolmogorov-Smirnov (K-S)  non-parametric two-sample test of the equality of continuous one-dimensional distributions. The K-S test computes the maximal distance between cumulative distribution functions and provides rejection thresholds for the null hypothesis that the compared samples are drawn from the same distribution. The question remains, how to find the distance function $D$ which produces networks that minimize the K-S statistic for centrality measure distributions.

Let $\delta(v_i,v_j)$ be the set of shortest paths between vertices $v_i$ and $v_j$ in the network $G$, and let $\delta_k(v_i,v_j)$ be the set of shortest paths between vertices $v_i$ and $v_j$ which pass through vertex $v_k$. Finally, let $\Delta(v_i,v_j)$ denote the length of the shortest path between vertices $v_i$ and $v_j$. A \emph{centrality measure} is a function $\mathcal{C}:V\rightarrow \mathbb{R}$ which assigns to each vertex a value representing the "importance" of the vertex in the network $G$. Four the most popular centrality measures include\cite{freeman1978centrality}:

\begin{itemize}
\item \emph{degree centrality} $C_D(v_i)=d(v_i)$ simply measures the number of vertices adjacent to the vertex $v_i$. The assumption here is that a vertex is important if it is directly connected to many vertices in the network.
\item \emph{betweenness centrality} $C_B(v_i)=\sum_{j,k \neq i} \left| \delta_i(v_j,v_k) \right|$ measures the number of shortest paths between any pair of vertices which pass through vertex $v_i$. This interpretation of importance highlights the influence of a vertex on communication pathways through the network.
\item \emph{closeness centrality} $C_C(v_i)= \frac{1}{\left| V \right|} \sum_j \Delta(v_i,v_j)$ measures the average distance from the vertex $v_i$ to all other vertices in the network. According to this definition, a vertex is important if it can quickly communicate with all remaining vertices in the network.
\item \emph{page rank centrality} $C_P(v_i)= \alpha \sum_{j:(v_j,v_i) \in E} \frac{C_P(v_j)}{d(v_j)} + \frac{1-\alpha}{n}$ measures the importance of a vertex as a recursive sum of importances of vertices adjacent to $v_i$. According to this definition, a vertex is important if it connects to other important vertices in the network.
\end{itemize}

Sometimes it is possible to ``guess'' the distance function $D$ given the description of the empirical network. Table~\ref{tab:distance.functions} presents a list of simple distance functions that can be used when re-creating empirical networks. Some of these distance functions are self-explanatory (like degree, betweenness, closeness, and page rank distances which are simply the expressions of preferential attachment to vertices with high values of these centrality measures). If the empirical network consists of vertices with attributes, euclidean distance can be used to generate local rankings of vertex priority. For the sake of simplicity we include only one- and two-dimensional euclidean distance. If vertices are described by numerical vectors, cosine distance can be utilized. The aggregate distance computes the distance on each pair of attribute values of compared vertices $v_i, v_j$, and then applies weights $w_k$ to distances computed on each attribute. 

\begin{table}
\centering
\begin{tabular}{llc}
\hline
name & formulation \\
\hline
random distance &  $D(v_i,v_j) \sim N(\mu,\sigma)$ \\
degree distance &  $D(v_i,v_j) = \frac{1}{C_D(v_j) + \epsilon}$ \\
betweenness distance &  $D(v_i,v_j) = \frac{1}{C_B(v_j) + \epsilon}$ \\
closeness distance &  $D(v_i,v_j) = \frac{1}{C_C(v_j) + \epsilon}$ \\
page rank distance & $D(v_i,v_j) = \frac{1}{C_P(v_j) + \epsilon}$ \\
euclidean 1-D distance &  $D(v_i,v_j) = \left| a^1_i - a^1_j \right| $ \\
euclidean 2-D distance &  $D(v_i,v_j) = \sqrt{(a^1_i - a^1_j)^2+ (a^2_i - a^2_j)^2} $ \\
cosine distance &  $D(v_i,v_j) = 1 - \frac{v_i \circ v_j}{\|v_i\| \cdot \| v_j \|}$ \\
aggregate distance &  $D(v_i,v_j) = \sum\limits_{k=1}^m w_k D(a^k_i,a^k_j)$ \\
linear regression distance &  $D(v_i,v_j) = W_{ij}\beta$, $W_{ij}=(a^1_i, \ldots, a^p_i, a^1_j, \ldots, a^p_j, \epsilon)$ \\
naive bayes classifier distance & $D(v_i,v_j) = \frac{P(C=1|W_{ij})}{P(C=0|W_{ij}) + \epsilon}$ \\
\end{tabular}
\caption{Distance functions}
\label{tab:distance.functions}
\end{table}

However, in most cases it is impossible to approximate the generative process of a network using a single, simple distance function. Given an empirical network, it is often precisely the aim of a researcher to deduce the guiding generative principle of a network. The main advantage of the Priority Rank model is its ability to derive the proper distance function $D$ from the existing network, which, if applied to the model, would generate the network most similar to the original one. The learning task can be defined as follows. Consider a network $G=\left< V,E \right>$, and in particular, consider an edge $(v_i,v_j) \in E$. For the Priority Rank model to re-generate this edge, it is expected the distance $D(v_i,v_j)$ to be minimized, and at the same time, the distance $D(v_i,v_k)$ should be maximized $\forall v_k : (v_i,v_k) \notin E$. The network $G$ provides the training data for a machine learning algorithm under the form of the adjacency matrix, which can be interpreted as a function:

\[
    A: V \times V \rightarrow \{0,1\} : \quad  A(v_i,v_j) = 
    \begin{cases}
        1,  & \text{if } (v_i, v_j) \in E \\
        0,  & \text{otherwise}
    \end{cases}
\]

The positive cases in the training set consist of tuples $(v^1_i, \ldots, v^m_i, v^1_j, \ldots, v^m_j)$ for all pairs of vertices $(v_i, v_j)$, which are adjacent in $G$, and the negative cases consist of tuples $(v^1_i, \ldots, v^m_i, v^1_k, \ldots, v^m_k)$ for all pairs of vertices $(v_i,v_k)$ which are not adjacent in $G$. The training set can be fed into a classification algorithm, such as logistic regression, naive Bayes classifier, or SVM, to find patterns in vertex attribute co-occurrences which influence the probability of edge's presence or absence. The model resulting from a classification algorithm can be interpreted as a condensed representation of the underlying principle of network formation. The last two distance functions presented in Table~\ref{tab:distance.functions} represent the learning procedure. The linear regression distance uses the least-squares method to fit the linear regression equation to the training set, and the naive Bayes classifier distance uses the well-known naive Bayes classifier to predict the probability of existence of an edge between two vertices. 

\section*{Results}
\label{sec:results}

The main result reported in this paper is the development of the attachment priority mechanism. In this section, we present the outcomes of conducted experiments. Their aim was to verify whether the priority attachment mechanism is capable of explaining the underlying network formation process. We have used four popular generative network models to produce synthetic networks, and we have collected 18 empirical networks from various domains to test the ability of the priority attachment to re-create these networks. The experimental protocol was as follows. Since we cannot guess which distance function will be able to best reproduce a given $G$, we have applied all distance functions presented in Table~\ref{tab:distance.functions}. Then, for the best three distance functions we have run the generation process 20 times, aggregating the results afterwards. Of course, most of the distance functions require values of attributes describing vertices. These values were not always available, for instance, in the case of synthetic networks generated from theoretical network models. In such cases, we have created synthetic attributes for each vertex, generating four attributes: one ordinal, one categorical, two continuous. These attributes were generated from four different distributions: normal, uniform, log-normal, and the exponential distribution. For empirical networks, we have generated synthetic attributes only when no vertex attributes were present in the data, otherwise we have used only the real features of vertices. To compare synthetic and empirical networks with networks generated by the Priority Rank model, we have tested the conformity of centrality measure distributions using the Kolmogorov-Smirnov two sample test. Recall that the null hypothesis of two sample K-S tests states that the compared samples are drawn from the same distribution. We reject the null hypothesis for p-values less than the significance level $\alpha=0.05$. In Tables~\ref{tab:random.network.results}~and~\ref{tab:empirical.network.results} we mark the results which pass the K-S test in boldface, i.e., instances where the null hypothesis holds.

\subsection*{Synthetic networks}
\label{subsec:artificial.networks}

We have tested the ability of the Priority Rank model to re-create networks using both synthetic networks obtained from the theoretical network models, and empirical networks representing various domains. Generative network models used in our experiments include the following:

\begin{itemize}
\item \emph{Erd\"os-R\'enyi random model}~\cite{erd6s1960evolution}: an empty network with $n=50$ vertices is created, and then, for each pair of vertices an edge is formed with the probability $p=0.4$.
\item \emph{Watts-Strogatz small world model}~\cite{watts1998collective}: initially, $n=50$ vertices are connected in a ring topology, with each vertex connecting to its $k=3$ neighbors, and then, each edge is randomly rewired with the probability $p=0.01$.
\item \emph{Albert-Barab\'asi scale free model}~\cite{barabasi1999emergence}: initial topology of the network consists of $n_0$ vertices forming a complete graph $K_{n_0}$, remaining vertices are added to the network sequentially until the desired number $n=50$ of vertices is reached, and each newly added vertex creates $k=3$ edges to existing vertices, choosing target vertices with the probability proportional to their degrees, hence the alternative name of the model: preferential attachment model. The regular, linear preferential attachment is achieved for $\alpha=1$ .
\item \emph{Leskovec forest fire model}~\cite{leskovec2005graphs}: vertices are added sequentially to the network, each out of $n=50$ new vertex provides $k$ edges to uniformly selected targets, and then adds more edges to direct neighbors of selected targets with the burning probability $p=0.3$, upon successful creation of a new edge the process continues recursively. The model produces networks of low density and relatively large diameter, but with average shortest path lengths significantly lower than in the case of small world networks. 
\item \emph{Dorogovtsev-Goltsev-Mendes hierarchical model}~\cite{PhysRevE.65.066122}: initially, the network consists of a single edge connecting two vertices, and in each step of the network growth an edge produces a new vertex which is immediately attached to both ends of the edge. Although the resulting network is not strictly a fractal, it has a strong hierarchical structure with power law distributions of vertex degree and local clustering coefficient. In addition, the network contains multiple loops and is far from a trivial tree-like structure.
\item \emph{Disassortative model}: an empty network with $n=100$ vertices is created and the following procedure is iteratively repeated. Vertices are ordered by their identifiers, the first 10 vertices create between 30 and 40 edges to randomly selected vertices from the range $\left< 50,100 \right>$, vertices from the range $\left< 10,50 \right>$  create up to 15 edges to random vertices, and vertices from the range $\left<51,100\right>$ create between 0 and 2 edges to random vertices. After each iteration self-loops and multiple edges are removed. The procedure continues until the degree assortativity coefficient falls below $-0.4$; this particular threshold has been set arbitrarily.
\end{itemize}

\begin{table}
\centering
\begin{tabular}{ll}
\hline
symbol & meaning \\
\hline
$p_D$ & p-value of the Kolmogorov-Smirnov test comparing the degree distributions of two networks \\
$p_B$ & p-value of the Kolmogorov-Smirnov test comparing the betweenness distributions of two networks \\
$p_C$ & p-value of the Kolmogorov-Smirnov test comparing the closeness distributions of two networks \\
$d$ & network diameter: the length of the longest shortest path in the network \\
$\rho$ & network density: the ratio of the number of existing edges to the maximum number of possible edges \\
$L$ & the average length of the shortest paths in the network \\
$\rho^2_{C_D}$ & Freeman's centralization of the degree distribution of the network \\
\hline
\end{tabular}
\caption{Metrics used to compare networks}
\label{tab:metrics}
\end{table}


\begin{table}[tbp]
\centering
\begin{tabular}{lccccccccccl}
    \hline
    network & $p_D$ & $p_B$ & $p_C$ & $\left| V \right|$ & $\left| E \right|$ & $d$ & $\rho$ & $L$ & $\rho^2_{C_D}$ \\
    \hline
    \multirow{2}{*}{random} & \multirow{2}{*}{\textbf{0.52}} & \multirow{2}{*}{\textbf{0.84}} & \multirow{2}{*}{\textbf{0.78}} & \multirow{2}{*}{50} & \multirow{2}{*}{1007} & 2.00 & 0.40 & 1.59 & 0.11 & original \\
    & & & & & & \textbf{2.17} & \textbf{0.42} & \textbf{1.59} & \textbf{0.12} & random distance \\
    \hline
    \multirow{2}{*}{small world} & \multirow{2}{*}{0.01} & \multirow{2}{*}{\textbf{0.05}} & \multirow{2}{*}{\textbf{0.05}} & \multirow{2}{*}{50} & \multirow{2}{*}{150} & 11.00 & 0.06 & 4.54 &  0.01 & original \\
    & & & & & & 14.33 & \textbf{0.06} & \textbf{4.84} &  0.05 & euclidean distance \\
    \hline
    \multirow{2}{*}{scale free} & \multirow{2}{*}{\textbf{0.37}} & \multirow{2}{*}{\textbf{0.08}} & \multirow{2}{*}{\textbf{0.05}} & \multirow{2}{*}{50} & \multirow{2}{*}{144} & 3.00 & 0.06 & 1.53 & 0.25 & original \\
    & & & & & & 4.18 & \textbf{0.06} & 1.83 & 0.38 & linear regression distance \\
    \hline
    \multirow{2}{*}{forest fire} & \multirow{2}{*}{\textbf{0.10}} & \multirow{2}{*}{\textbf{0.60}} & \multirow{2}{*}{\textbf{0.20}} & \multirow{2}{*}{50} & \multirow{2}{*}{93} & 7.00 & 0.04 & 2.16 & 0.07 & original \\
    & & & & & & \textbf{7.09} & 0.05 & \textbf{2.37} & 0.08 & euclidean distance \\
    \hline
\multirow{2}{*}{hierarchical} & \multirow{2}{*}{\textbf{0.11}} & \multirow{2}{*}{\textbf{0.71}} & \multirow{2}{*}{0.00} & \multirow{2}{*}{123} & \multirow{2}{*}{246} & 5.00 & 0.02 & 1.70 &  0.12 & original \\
& & & & & & 6.00 & \textbf{0.02} & 2.08 &  0.16 & degree distance \\
    \hline
\multirow{2}{*}{disassortative} & \multirow{2}{*}{\textbf{0.28}} & \multirow{2}{*}{\textbf{0.37}} & \multirow{2}{*}{0.00} & \multirow{2}{*}{100} & \multirow{2}{*}{510} & 8.00 & 0.05 & 2.85 &  0.11 & original \\
& & & & & & 7.00 & \textbf{0.05} & \textbf{2.88} & 0.25 & degree distance \\
    \hline
\end{tabular}
\caption{Re-creation of synthetic networks. Results that pass the K-S test as well as scalar network descriptors such as diameter $d$ or density $\rho$ that are regenerated within $\pm10\%$ margin of the original value are marked with bold font}
\label{tab:random.network.results}
\end{table}

Metrics used to compare synthetic networks and networks generated by the Priority Rank model are presented in Table~\ref{tab:metrics}. Synthetic network statistics and the corresponding statistics of networks generated by the Priority Rank model are shown in Table~\ref{tab:random.network.results}. For distributions of degree, betweenness, and closeness, a standard two sample Kolmogorov-Smirnov test at level $\alpha=0.05$ has been performed and cases when the null hypothesis cannot be rejected are marked with boldface, i.e., the null hypothesis states that the compared samples come from the same distribution. Please note, however, that even if the re-created value does not fulfill this condition, it is rarely significantly different from the original since the original network descriptor is very small.

\subsubsection*{Erd\"os-R\'enyi random network}

The Priority Rank model is able to re-create Erd\"os-R\'enyi random network very well. As can be seen in Table~\ref{tab:random.network.results}, the network generated by the Priority Rank model reproduces all three centrality distributions of degree, betweenness, and closeness. It also retains the average shortest path length, and Freeman's centralization of degree distribution. The only network statistic which is not duplicated is diameter. Overall, edges are created randomly in the Erd\"os-R\'enyi model, so random distance function works very well because vertices have no preference for other vertices. 

\subsubsection*{Watts-Strogatz small world network}

The initial topology of the ring is formed in one dimension. None of the considered distance functions was able to re-create the degree distribution, although the euclidean distance function managed to generate networks with similar betweenness and closeness distributions. The standard deviation of the K-S statistic is quite large, which weakens the result. The euclidean distance function used two synthetic attributes generated from exponential and normal distributions. The failure to re-generate the degree distribution is caused by the fact that Priority Rank does not have anything akin to random edge re-wiring, which is essential for the small-world model. The ring structure can be very easily reproduced. However, the presence of a few randomly re-wired edges, which reduce average path lengths in the network, is very difficult to mimic using any distance function. Apart from this deficiency, the Priority Rank model generates networks, which are very similar to the original small world network in terms of network diameter, density, and average shortest path lengths, these characteristics are re-created almost flawlessly. 

As a matter of fact, Priority Rank is capable of producing small world networks. The main issue is with reproducing the topology of a particular implementation of the Watts-Strogatz model. The original Watts-Strogatz model places all vertices on the circumference of a circle and attaches neighboring nodes. Thus, the proximity of vertices is defined by the Euclidean distance in two dimensions, with vertices artificially placed on a circle. This is a highly contrived scenario which bares little resemblance to any practical application of the model. On the other hand, we have assumed in our experiment that each vertex has a single numerical attribute. The proximity of vertices is defined by the one dimensional distance in the space of this attribute. For instance, one can interpret this attribute as yearly income. The distance function reflects the tendency of people to socialize within levels of financial hierarchy. Another example could be the tendency of people to form social ties with people of similar age, i.e., age is the attribute. Any scenario when the social network is driven by strong homophily can be modeled using our distance function, provided that the driver of homophily can be expressed by a single numerical attribute. For $n=100$ vertices, neighborhood size of $m=4$ and the probability of random edge rewiring $p=5\%$ the Priority Rank model produces the network with the global clustering coefficient (also known as  transitivity) of $\mathit{gcc}=0.41$. The native function \texttt{igraph::sample\_smallworld()} from the \texttt{igraph\textbackslash R} package with the same parameters delivers a network with $\mathit{gcc}=0.47$. It is, in turn, far away from the network produced by Priority Rank using another degree distance function (the output network is scale free): $\mathit{gcc}=0.10$ as well as from the network provided by the random distance function (that mimics the Erd\"os-R\'enyi network): $\mathit{gcc}=0.06$. In our opinion, the Priority Rank model still produces valid small world networks defined as networks, which have disproportionately large global clustering coefficient and relatively small average shortest path length.

\subsubsection*{Albert-Barab\'asi scale free network}

Priority Rank successfully re-creates degree and betweenness distributions, while struggling to preserve the distribution of the closeness centrality (the linear regression distance function barely manages to pass the K-S test). The Priority Rank model also builds networks with larger diameters than the original network, but the density and the centralization of the degree distribution are very close to original values. We also note that the networks generated based on Priority Rank, in which average shortest paths are slightly longer than in the original Albert-Barab\'asi model. Nevertheless, we conclude that synthetic scale-free networks can be mimicked by the Priority Rank model sufficiently.

\subsubsection*{Leskovec forest fire network}

The Priority Rank model can re-create forest fire networks very precisely. The best results are obtained when using euclidean distance based on one attribute with values drawn randomly from the normal distribution. Such distance function delivers networks with similar centrality measure distributions (degree, betweenness, closeness), and with very similar diameter, density, average shortest path lengths and degree distribution. 

\subsubsection*{Dorogovtsev-Goltsev-Mendes hierarchical network}

The Priority Rank model is capable of successful generating hierarchical network structures, such as pseudo-fractal topologies of the Dorogovtsev-Goltsev-Mendes network. The only centrality measure that has not been reproduced is the closeness. Network density is reproduced exactly whereas the remaining descriptors are rather close to their original values.

\subsubsection*{Disassortative network}

As can be expected, the degree distance function was the best choice for re-creating disassortative networks. Distributions of degree and betweenness are re-created very well together with density and the average shortest path length. The diameter of the network is almost reproduced, and the only network descriptor which significantly differs from the original is Freeman's degree centralization.

\vspace{0.7cm}

Overall, the Priority Rank model can easily mimic synthetic networks produced by popular network generative models. A simple substitution of the distance function allows the Priority Rank model to produce instances of random networks, small world, scale-free, forest fire, hierarchical, and disassortative networks. In addition, one can easily introduce different variations of these generative models by modifying the distance function used to compute local priority rankings. This flexibility and ability to generalize multiple models is a unique and practical feature of the Priority Rank model.

\subsection*{Empirical networks}
\label{subsec:empirical.networks}

\begin{table}
\centering
\begin{tabular}{lllcc}
\hline
type & name & description \\
\hline
\multirow{5}{*}{biol-chem} 
& \emph{american bison} & Dominance relations among a group of American bison bulls  \\ 
& \emph{bighorn sheep} &  Dominance interactions among a group of female bighorn sheep \\
& \emph{C.elegans} & Neural connections of the C.elegans nematode \\
& \emph{mouse visual cortex} & Neuron interactions in the mouse primary visual cortex \\
& \emph{enzyme 108} & Enzyme interaction network  \\
& \emph{cage5} & Cage model of DNA electrophoresis  \\
\hline
\multirow{8}{*}{social} 
& \emph{political books} & Co-purchase of books about U.S. politics on Amazon \\ 
& \emph{primary school} & Contacts among students and teachers at a primary school in Lyon \\
& \emph{vickers 7th graders} & Friendships among seventh grade students in Victoria, Australia \\
& \emph{freeman researchers} & E-mail messages exchanged between researchers \\
& \emph{terrorists} & Social associations of 9/11 hijackers \\
& \emph{karate club} & Friendships among members of a university karate club \\
& \emph{illinois high school} & Friendships among male students in a small high school in Illinois \\
& \emph{marseille high school} & Contacts among students in a high school in Marseilles \\
\hline
\multirow{4}{*}{misc} 
& \emph{CAG} & Integer matrices used in characteristic polynomials computations \\ 
& \emph{power network} & Symmetric structure of the standard test of the power system network \\
& \emph{football} & Foosball player contracts between countries \\
& \emph{st.marks ecosystem} & Carbon-flow among species in St.Marks National Wildlife Refuge \\
\hline
\end{tabular}
\caption{Empirical networks used in the experiments}
\label{tab:real.world}
\end{table}

Apart from popular generative network models, the Priority Rank model can generate networks with topologies not available through traditional models if provided with a suitable custom distance function. However, the most interesting and valuable property of the Priority Rank model is its ability to learn the generative processes of empirical networks and to generate multiple instances of these networks. 
The experimental evaluation of this feature is carried out on networks listed in Table~\ref{tab:real.world} and in the supplementary material. They are provided by The Colorado Index of Complex Networks~\cite{icon} and The Network Repository~\cite{nr}. Metrics used to compare empirical networks and networks generated by the Priority Rank model are presented in Table~\ref{tab:metrics}. Statistics of original networks along with corresponding ones for the networks generated by Priority Rank are shown in Table~\ref{tab:empirical.network.results}.

\begin{table}[tbp]
\centering
\begin{tabular}{lccccccccccl}
    \hline
    network & $p_D$ & $p_B$ & $p_C$ & $\left| V \right|$ & $\left| E \right|$ & $d$ & $\rho$ & $L$ & $r$ & $\rho^2_{C_D}$ \\
    \hline
    \multirow{2}{*}{american bisons} & \multirow{2}{*}{\textbf{0.23}} & \multirow{2}{*}{\textbf{0.27}} & \multirow{2}{*}{\textbf{0.67}} & \multirow{2}{*}{26} & \multirow{2}{*}{313} & 4 & 0.46  & 1.57 & 0.59 & 0.31 & original        \\
    & & & & & & 3 & \textbf{0.44} & \textbf{1.59} & 0.42 & 0.19 & random distance \\
    \hline
    \multirow{2}{*}{bighorn sheep} & \multirow{2}{*}{\textbf{0.33}} & \multirow{2}{*}{\textbf{0.68}} & \multirow{2}{*}{\textbf{0.17}} & \multirow{2}{*}{28} & \multirow{2}{*}{250} & 5 & 0.32  & 1.90 & 0.12 & 0.16 & original        \\
    & & & & & & 4 & \textbf{0.34} & \textbf{1.86} & 0.31 & 0.31 & page rank distance \\
    \hline
    \multirow{2}{*}{C.elegans} & \multirow{2}{*}{\textbf{0.35}} & \multirow{2}{*}{\textbf{0.35}} & \multirow{2}{*}{0.00} & \multirow{2}{*}{279} & \multirow{2}{*}{2993} & 7 & 0.04  & 2.88 & 0.47 & 0.21 & original        \\
    & & & & & & \textbf{7} & \textbf{0.04} & \textbf{2.83} & 0.04 & 0.13 & closeness distance \\
    \hline
    \multirow{2}{*}{mouse visual cortex} & \multirow{2}{*}{\textbf{0.95}} & \multirow{2}{*}{\textbf{0.95}} & \multirow{2}{*}{0.00} & \multirow{2}{*}{29} & \multirow{2}{*}{44} & 2 & 0.05  & 1.08 & 0.00 & 0.11 & original        \\
& & & & & & 4 & 0.06 & 1.76 & 0.12 & 0.18 & page rank distance \\
    \hline
\multirow{2}{*}{enzyme 108} & \multirow{2}{*}{\textbf{0.05}} & \multirow{2}{*}{\textbf{0.54}} & \multirow{2}{*}{\textbf{0.24}} & \multirow{2}{*}{38} & \multirow{2}{*}{164} & 15 & 0.22  & 5.85 & 1.00 & 0.10 & original        \\
& & & & & & \textbf{14} & 0.26 & \textbf{5.39} & 0.58 & 0.06 & euclidean distance \\
    \hline
\multirow{2}{*}{cage5} & \multirow{2}{*}{\textbf{0.77}} & \multirow{2}{*}{\textbf{0.89}} & \multirow{2}{*}{0.00} & \multirow{2}{*}{72} & \multirow{2}{*}{940} & 6 & 0.18  & 2.22 & 0.56 & 0.31 & original        \\
& & & & & & 7 & \textbf{0.18} & \textbf{2.10} & 0.23 & 0.40 & degree distance \\
    \hline
\multirow{2}{*}{political books} & \multirow{2}{*}{\textbf{0.31}} & \multirow{2}{*}{\textbf{0.73}} & \multirow{2}{*}{0.00} & \multirow{2}{*}{105} & \multirow{2}{*}{441} & 8 & 0.04  & 2.92 & 0.00 & 0.08 & original        \\
& & & & & & 7 & \textbf{0.04} & \textbf{3.00} & 0.14 & 0.12 & euclidean distance \\
    \hline
\multirow{2}{*}{primary school} & \multirow{2}{*}{\textbf{0.81}} & \multirow{2}{*}{0.00} & \multirow{2}{*}{0.00} & \multirow{2}{*}{242} & \multirow{2}{*}{8317} & 4 & 0.29  & 1.77 & 0.00 & 0.14 & original        \\
& & & & & & \textbf{4} & \textbf{0.30} & 2.03 & 0.30 & 0.22 & euclidean distance \\
    \hline
\multirow{2}{*}{vickers 7th graders} & \multirow{2}{*}{\textbf{0.78}} & \multirow{2}{*}{\textbf{0.78}} & \multirow{2}{*}{\textbf{0.37}} & \multirow{2}{*}{29} & \multirow{2}{*}{376} & 3 & 0.45  & 1.60 & 0.67 & 0.32 & original        \\
& & & & & & \textbf{3} & \textbf{0.48} & \textbf{1.59} & 0.42 & \textbf{0.32} & closeness distance \\
    \hline
\multirow{2}{*}{freeman researchers} & \multirow{2}{*}{\textbf{0.83}} & \multirow{2}{*}{\textbf{0.09}} & \multirow{2}{*}{\textbf{0.96}} & \multirow{2}{*}{32} & \multirow{2}{*}{460} & 3 & 0.45  & 1.56 & 0.79 & 0.55 & original        \\
& & & & & & \textbf{3} & \textbf{0.45} & \textbf{1.58} & 0.51 & \textbf{0.52} & betweenness distance \\
    \hline
\multirow{2}{*}{terrorists} & \multirow{2}{*}{\textbf{0.68}} & \multirow{2}{*}{\textbf{0.82}} & \multirow{2}{*}{0.00} & \multirow{2}{*}{62} & \multirow{2}{*}{304} & 5 & 0.16  & 2.95 & 1.00 & 0.28 & original        \\
& & & & & & 8 & 0.14 & \textbf{2.84} & 0.07 & \textbf{0.28} & degree distance \\
    \hline
\multirow{2}{*}{karate club} & \multirow{2}{*}{\textbf{0.86}} & \multirow{2}{*}{\textbf{0.06}} & \multirow{2}{*}{0.00} & \multirow{2}{*}{34} & \multirow{2}{*}{78} & 3 & 0.13  & 1.27 & 0.00 & 0.19 & original        \\
& & & & & & 4 & \textbf{0.14} & 1.68 & 0.04 & \textbf{0.19} & closeness distance \\
    \hline
    \multirow{2}{*}{illinois high school} & \multirow{2}{*}{\textbf{0.53}} & \multirow{2}{*}{\textbf{0.61}} & \multirow{2}{*}{\textbf{0.06}} & \multirow{2}{*}{70} & \multirow{2}{*}{336} & 12 & 0.07  & 3.97 & 0.50 & 0.09 & original        \\
    & & & & & & \textbf{11} & 0.08 & \textbf{3.91} & 0.35 & 0.06 & aggregate distance \\
    \hline
    \multirow{2}{*}{marseille high school} & \multirow{2}{*}{\textbf{0.10}} & \multirow{2}{*}{0.02} & \multirow{2}{*}{0.00} & \multirow{2}{*}{126} & \multirow{2}{*}{1710} & 6 & 0.11  & 2.03 & 0.00 & 0.11 & original        \\
    & & & & & & \textbf{6} & \textbf{0.11} & 2.40 & 0.10 & 0.31 & linear regression distance \\
    \hline
    \multirow{2}{*}{CAG} & \multirow{2}{*}{\textbf{0.89}} & \multirow{2}{*}{\textbf{0.09}} & \multirow{2}{*}{0.00} & \multirow{2}{*}{72} & \multirow{2}{*}{940} & 6 & 0.18  & 2.22 & 0.56 & 0.31 & original        \\
& & & & & & 7 & 0.11 & 2.55 & 0.12 & 0.23 & betweenness distance \\
    \hline
    \multirow{2}{*}{power network} & \multirow{2}{*}{\textbf{0.75}} & \multirow{2}{*}{\textbf{0.91}} & \multirow{2}{*}{\textbf{0.09}} & \multirow{2}{*}{39} & \multirow{2}{*}{46} & 11 & 0.06  & 4.39 & 0.00 & 0.04 & original        \\
    & & & & & & 13 & 0.08 & \textbf{4.77} & \textbf{0.00} & 0.03 & random distance \\
    \hline
    \multirow{2}{*}{football} & \multirow{2}{*}{\textbf{0.98}} & \multirow{2}{*}{\textbf{0.32}} & \multirow{2}{*}{0.00} & \multirow{2}{*}{35} & \multirow{2}{*}{118} & 3 & 0.10  & 1.30 & 0.00 & 0.19 & original        \\
    & & & & & & 4 & \textbf{0.11} & 1.88 & 0.13 & 0.31 & degree distance \\
    \hline
    \multirow{2}{*}{st.marks ecosystem} & \multirow{2}{*}{\textbf{0.89}} & \multirow{2}{*}{\textbf{0.09}} & \multirow{2}{*}{0.00} & \multirow{2}{*}{54} & \multirow{2}{*}{353} & 7 & 0.12  & 2.87 & 0.02 & 0.34 & original        \\
    & & & & & & \textbf{7} & \textbf{0.11} & 2.55 & 0.12 & 0.23 & betweenness distance \\
    \hline
\end{tabular}
\caption{Re-creation of empirical networks from Table~\ref{tab:real.world}. Results that pass the K-S test and scalar descriptors from within $\pm 10\%$ margin of the source network are in bold}
\label{tab:empirical.network.results}
\end{table}

\begin{itemize}
    \item \textbf{American bisons}: The Priority Rank model correctly captures the underlying network structure and can re-create the network to a sufficient extent. All centrality measure distributions are retained in the generated networks, and these networks exhibit densities, and average shortest path lengths similar to the original network. The Priority Rank model slightly underestimates the diameter and the centralization of the degree distribution. Since the original network resembles the Erd\"os-R\'enyi random network, no wonder that the random distance best matches recreation process. Since Priority Rank is capable of re-creating the \emph{American bison} network, this means that one could generate multiple instances of this network and belief that the generated instances reflect the same principle guiding the formation of the original network. These new networks could be interpreted as observations of another herd of bisons, or snapshots of the same group for different periods. This may be very useful for simulation purposes while analyzing transmission of a disease between individual animals living in the wild. Network instances generated with Priority Rank could be used as multiple alternative scenarios.

    \item \textbf{Bighorn sheep}: The Priority Rank model very convincingly re-creates the topology of the original network maintaining distributions of all centrality measures, density and the average shortest path lengths. The generated networks have slightly smaller diameters than the original one, and the reciprocity is over-estimated. This experiment, however, supports our general claim that the Priority Rank model can discover the latent generative principle of an empirical network. The page rank distance aggregates the importance of all animals in the herd. Intuitively page rank reflects the true importance of animals given the history of dominance relationships. For the sake of brevity, we report only on the best fitting distance function for each network. Nevertheless, we have analyzed several different distance functions for each network. In this case, very similar results were obtained for two distance functions, which both made use of the age attribute. Naive Bayes classifier distance and linear regression distance functions are machine learning algorithms which map the relationship between two animals' ages onto the probability of the existence of the dominance relationship between these animals. These two distance functions were close competitors of the page rank distance function. Thus, the interpretation of these distance functions allows us to extract a general principle of this particular network formation. 
    \item \textbf{C.elegans}: The Priority Rank model generates networks which are very similar in terms of degree and betweenness distributions to the \emph{C.elegans} network, but we were unable to re-create a similar distribution of the closeness centrality measure (this could be due to a large number of vertices in the network). Also the distance, the density and the average path lengths are faithfully re-created with Priority Rank. One missing characteristic of the original network is the reciprocity. Our model cannot capture this feature using the closeness distance function. We note, however, that adding reciprocity to the distance function is straightforward. It is sufficient to include a component that would diminish the estimated distance between vertices $v_i$ and $v_j$, if an edge $(v_j,v_i)$ is already present in the network.
    
    \item \textbf{Mouse visual cortex}: This small network is very difficult to re-create, most probably due to the fact that its generative process may be complex. Unfortunately, the source network does not contain any additional attributes and we have to rely only on topological features of the network. In this case, the best result is obtained for the page rank distance which creates networks similar to the scale free model. The Priority Rank model is able to produce networks with very similar degree and betweenness distributions.

    \item \textbf{Enzyme 108}: The Priority Rank model using euclidean distance can re-generate all centrality measure distributions and obtain very similar values of diameter, density, and the average shortest paths length. The euclidean distance measure uses two synthetic attributes drawn from the uniform distribution. A distance function tends to produce networks similar to the small world model, and the \emph{enzyme 108} network definitely belongs to this network family. This small-worldliness of \emph{enzyme 108} is best manifested in the density and the average shortest path length, which are the largest among all analyzed networks. Interestingly, although Priority Rank had problems re-creating the synthetic small-world network, it manages to approximate the real world example of a small-world network very faithfully. Look at \emph{Illinois high school} to see the same Priority Rank behavior. This might indicate that the original small-world model proposed by Watts and Strogatz over-simplifies the reality and the priority attachment based on the similarity of attributes is a better representation of real network formation phenomena.

    \item \textbf{Cage5}: The degree distance function reproduces distributions of degree and betweenness, but cannot re-create the original distribution of the closeness centrality. As for the topology of the generated network, the Priority Rank model generates very similar networks in terms of diameter, density, average shortest path length and degree centralization. Our model underestimates the reciprocity of the network, most probably due to the fact that the degree distance function alone cannot account for the increased probability of reciprocal relationships.

    \item \textbf{Political books}: This network is easily re-created with the Priority Rank model using the euclidean distance function based on a single discrete attribute, which reduces the function to a simple binary flag comparison. This result agrees very well with our intuition. One can interpret the synthetic discrete attribute as an indicator of a broad book category, and two books are purchased together if they belong to the same category. The Priority Rank model re-creates degree and betweenness distributions, but fails to retain the original distribution of closeness. Also the diameter, the density, the average shortest path length, and the centralization of degree are very similar to the original network. The difference in the reciprocity estimation results from lack of direction of edges in the original network. The Priority Rank model produces inherently directed networks. This example illustrates well the ability of the Priority Rank model to discover the latent process of network formation and provide simple, interpretable explanations of the underlying network structure.
    
    \item \textbf{Primary school}: The main challenge that this network poses is the density of social interactions and very short average shortest path lengths. The Priority Rank model re-creates this network reasonably well using the euclidean distance function on a single discrete attribute. Similarly to the \emph{political books} network, a simple comparison of a discrete attribute is sufficient to produce a good approximation of the original network. Although the Priority Rank model cannot reproduce original distributions of betweenness and centrality, the degree distribution is very well preserved. The remaining topological characteristics of the network are also retained, except for the reciprocity --- for the same reasons as for \emph{political books} network. The interpretation of the distance function is similar to the \emph{political books} network, namely, the discrete attribute serves as a label of a coherent social group and students have a strong preference to create relationships within the social group they belong to.

    \item \textbf{Vickers 7th graders}: The Priority Rank model re-creates this network perfectly using the closeness distance, where the closeness of the vertex is computed on incoming edges. With the exception of the reciprocity (slightly underestimated), all the remaining network profile is re-created precisely. One reason for this result is the fact that multiple types of edges have been flattened to a single layer in the original network, making the guessing of edge existence a bit easier. Incoming closeness seems to be a very reasonable explanation for edge presence. Popular students who have been nominated by many peers have indeed high value of inbound closeness centrality. Using this measure to generate ranking lists in the Priority Rank model yields good approximations of the original network. Again, we see how the Priority Rank model provides a simple and interpretable explanation of the latent network generation process.

    \item \textbf{Freeman researchers}: An interesting feature of this network is the fact that it represents communication patterns in the pre-internet era. As expected, the network is characterized by very high reciprocity, low diameter and average shortest path length. The best distance function uses betweenness centrality. Our model not only re-generates identical distributions of centrality measures, but also captures all the remaining network statistics. Again, only the reciprocity has not been matched, due to the fact that the betweenness distance function is not capable of taking the existence of an edge into consideration. Since the network represents the flow of communication between people, it is not surprising that the best distance function uses the betweenness centrality. It ranks network vertices according to their impact on the communication pathways through the network. We find this result to be a strong indication that the Priority Rank model really can provide a viable explanation for the phenomenon driving the network formation. 

    \item \textbf{9/11 terrorists}: Similarly to the network of researchers discussed above, the Priority Rank model re-creates the 9/11 terrorists network precisely. Although the distribution of closeness centrality is not preserved, the remaining network characteristics are almost identical to the original network. The choice of the distance function is also obvious. The network has been gathered \emph{post hoc} in a way to outline the associations of a selected group of terrorists, and, as the result, these terrorists possess prominent degrees in the network. Again, the interpretation of the distance function is straightforward and the result supports our claim about the explanatory power of the Priority Rank model.

    \item \textbf{Karate club}: Priority Rank is capable of re-creating the structures of the original network, with the notable exception of the closeness distribution and the overestimated average shortest path length. The best distance function and perfect network explanation is the closeness distance. Recall that the network represents social contacts within the karate club community at a university. The community has split in half due to a conflict between the two principal members, the administrator and instructor. Almost all of the members of the original club have links to one of the leaders of the club. Thus, one may conclude that the members had a strong preference to prioritize social contacts to these leaders, who are the two vertices with the shortest average paths to all the remaining vertices.

    \item \textbf{Illinois high school}: The most distinguishing feature of this network is its very large diameter and relatively high average shortest path length. Theoretically, this network tends to follow the small world network model of Watts and Strogatz. The Priority Rank model re-generates this network almost perfectly, retaining the distributions of centrality measures and all other network descriptors. As expected, the best fitting is obtained for the aggregate distance function. This function compares the values of two attributes randomly drawn from the normal distribution. As a result, vertices with many similar attribute values tend to form communities (small worlds), with very few edges interconnecting these communities, hence the large network diameter. One may conclude that the evolution of these types of small world networks is primarily driven by the homogeneity of vertices constituting communities. We find it surprising that the Priority Rank model struggled to re-create the synthetic small world network, whereas it re-creates real world examples of small world networks very convincingly. In our opinion, this indicates that the assumptions behind the small world model of Watts and Strogatz are not fully supported by empirical networks.

    \item \textbf{Marseille high school}: Surprisingly, the Priority Rank model has struggled to reconstruct this network, not being able to re-create distributions of betweenness and closeness. However, despite the problem with retaining the original centrality measure distributions, the networks generated by Priority Rank are very similar w.r.t. the remaining network characteristics. The best distance function is linear regression, which indicates that the attributes of vertices (class, sex) are crucial for inferring the existence of relationships. This is even more encouraging than previous examples where networks have been re-created using synthetic attributes. The linear regression distance function can be easily interpreted. It even provides quantitative estimations of the latent network generative process by means of regression coefficients. 

    \item \textbf{CAG}: The main focus of this experiment was to check whether Priority Rank is capable of capturing the generative process of highly atypical network induced by integer matrices used in computations of characteristic polynomials. Our model re-creates degree and betweenness distributions, as well as most of the descriptors of the original network, except reciprocity. Since the best fitting function is the degree distance, we are tempted to conclude that the CAG network is generated primarily by the preferential attachment mechanism.

    \item \textbf{Power network}: This dataset representing the structure of power stations connections is challenging due to its very low density, large diameter and lack of degree centralization. Despite this, the Priority Rank model successfully reproduces the network, preserving all the distributions and measures. The best results are obtained for the random distance function, suggesting a haphazard setup of the network. 

    \item \textbf{Football}: Priority Rank produces networks very similar to the original network, with slightly over-estimated average shortest path length and degree centralization. The former is probably caused by the greater diameters than the original network. As expected, the best fitting function is the degree distance. It reflects the socio-economic reality of the process underlying the network formation: poorer countries are exporting best players to a few countries where very wealthy football clubs reside. Here we can see that the Priority Rank model provides an interpretation of the latent network formation mechanism, even if the mechanism involves complex socio-economical processes. 

    \item \textbf{St.Marks ecosystem}: In this experiment, we were particularly interested in discovering the underlying generative process of network formation. We can see that the Priority Rank model re-generates this network very accurately, preserving distributions of degree and betweenness and producing very similar diameters, densities, and average shortest path lengths. The betweenness distance turned out to be the best function. It is not surprising given the fact that the network focuses on modeling flows within the ecosystem. Again, we believe that the Priority Rank model discovers the main generative process of this network formation. 
\end{itemize}

\section*{Discussion}
\label{sec:discussion}

In this paper, we have developed the priority attachment, a plausible principle of network formation. We have shown that priority attachment can generalize previously proposed mechanisms of network formation, such as the small world phenomenon or preferential attachment. The second major contribution of this paper is the introduction of the Priority Rank model, a comprehensive generative network model which utilizes the priority attachment principle. The Priority Rank model is capable of mimicking both synthetic networks produced by popular generative network models, as well as re-creating empirical networks. The priority attachment mechanism requires a distance function to compute local rankings for each vertex. The interpretation of the distance function allows us to infer the properties of generated networks in case of synthetic networks. More importantly, it captures the nature of the generative process imprinting formation of empirical networks. In our opinion this is the most valuable and handy feature of Priority Rank. The ability to discover the guiding principle of network formation facilitates generation of  multiple similar but slightly different realizations for a given network, resulting in the whole population of networks. The stochastic nature of the priority attachment mechanism provides a small degree of randomization required to introduce variance into such population of networks, while preserving distributions of centrality measures and network profile at the same time. 

Additionally, the Priority Rank model and a suitable distance function can provide a network generation principle that better matches real networks than the other well-known generative models. The \emph{Enzyme 108} and \emph{Illinois high school} networks are quite typical examples of small-world networks that are better represented by the Priority Rank model than by the Watts-Strogatz model \cite{watts1998collective}. 

Many different applications of the Priority Rank model can be enumerated. Apart from the obvious substitution of multiple generative network models by a single model, the Priority Rank model can be used for A/B testing of networks. Using our model, series of networks with a given profile may be generated. As a result, statistical inference on networks is possible owing to presence of many network instances belonging to the same family as the original one, which represents only a single data point. Priority Rank facilitates the simulation of various scenarios for network formation and growth. Given a model of network formation principle under the form of the distance function, we can predict evolution of networks.  

Priority Rank is an early implementation of the priority attachment principle, which is capable of creating or re-creating networks of varying topologies, but it should not be regarded as a final model. In fact, in the next section we outline several limitations of the current model and we sketch possible solutions. But despite some deficiencies of the Priority Rank model we see strong evidence in favor of our main claim, namely, that priority attachment is a generalization of previously proposed network formation principles and can successfully be used to explain the emergence of many network topologies.

\subsection*{Limitations}
\label{subsec:limitations}

The main focus of this paper is the introduction of the priority attachment, an important and useful mechanism of network formation. Priority Rank, on the other hand, is a generative model which uses the priority attachment when generating networks. As such, Priority Rank has certain limitations which are unavoidable, since every network model has to make a compromise between accuracy and simplicity. Firstly, we note that the Priority Rank model does not always re-create empirical networks fully. It seems that the closeness distribution is notoriously difficult to reproduce. Also, we note the lack of a mechanism which would allow to easily account for the reciprocity. Of course, this can be introduced into the model by modifying the distance function to put more weight to existing edges. 

Another shortcoming of the current model is the fact, that while trying to re-create an empirical network, the search for the suitable distance function is performed via brute force approach. As of now, we are searching through the parameter grid of a fixed set of template distance functions and finding the distance function which best reproduces the input network. This is both time consuming and ineffective, as the search space is limited. Possible solution to this shortcoming would be to replace currently used distance functions by a single universal distance function discovery module. We believe that for larger networks, where adjacency matrices contain sufficient amount of information, the application of neural networks to form distance functions is a viable direction of future research. As a matter of fact, we are currently working on a deep neural network architecture that could extract several layers of topological features and learn the distance function.

Finally, we note that the amount of randomness in the Priority Rank model is limited, which is most pronounced in the inability of the model to re-create small world networks (at least the networks produced by the Watts-Strogatz model). There are two places where randomness can be introduced into the Priority Rank model: either the distance function can produce random results for a subset of vertices, or one can modify the rank probabilities used by the model to select target vertices from local vertex rankings. The latter can be achieved, e.g., by flattening the distribution of probabilities of picking vertices at given positions in the ranking. This issue, will be another subject of future research.

\bibliography{priorityrank}

\section*{Acknowledgments}

We wish to thank F.Menczer, S.Fortunato, and A.Flammini for their valuable remarks. 

This work was partially supported by the National Science Centre, Poland, projects no. 2016/23/B/ST6/03962, 2016/21/B/ST6/01463 and 2016/21/D/ST6/02948; European Union’s Horizon 2020 research and innovation program under the Marie Skłodowska-Curie grant agreement No. 691152 (RENOIR); the Polish Ministry of Science and Higher Education fund for supporting internationally co-financed projects in 2016-2019 no. 3628/H2020/2016/2.

The authors declare that there is no conflict of interest regarding the publication of this paper.

\section*{Additional information}

PriorityRank online generator is available at \url{https://priorityattachment.ml}

\section*{Author information}

\subsection*{Affiliations}

\hspace{0.5cm} \textbf{Institute of Computing Science, Poznań University of Technology, Poland}

Mikołaj Morzy, Grzegorz Miebs, Arkadiusz Rusin

\textbf{Faculty of Computer Science \& Management, Wroclaw University of Science and Technology, Poland}

Tomasz Kajdanowicz, Przemysław Kazienko

\subsection*{Author contributions statement}

M.M., T.K., and P.K. devised the idea and designed experiments. A.R. constructed the prototype and conducted initial experiments on real-world networks. G.M. extended the prototype and conducted follow-up experiments on real-world data. M.M., T.K., P.K., and G.M. analyzed the data and experimental results. M.M., T.K., and P.K. edited the manuscript.

\subsection*{Competing interests}

The authors declare no competing financial interests.

\subsection*{Corresponding author}

Correspondence to Mikołaj Morzy.

\end{document}